\documentclass[aps,prl,twocolumn,preprintnumbers,groupedaddress,superscriptaddress,floatfix,tightenlines,reprint]{revtex4-1}
\usepackage{mathrsfs,natbib}
\usepackage{graphics,epsfig,color, graphicx}
\usepackage{verbatim}
\usepackage{bm}

\usepackage{graphicx,epsf,amssymb,bbm,amsbsy,amsfonts,amssymb,amsmath}
\usepackage{hyperref}

\hypersetup{
    colorlinks=true,       
    linkcolor=red,          
    citecolor=blue,        
    filecolor=magenta,      
    urlcolor=blue           
}

\usepackage{amsmath,amssymb,multirow,graphicx}
\usepackage{centernot}
\usepackage{color}
\allowdisplaybreaks[1]
\usepackage{tikz}
 \tikzset{node distance=2cm, auto}
\usepackage{bbm} 
\usepackage{youngtab}

\renewcommand{\Im}{\text{Im }}

\def\Im{\text{Im}}

\def\tr{\text{tr}}

\def\bar{\overline}

\def\Z{{\mathbb Z}}
\def\R{{\mathbb R}}
\def\coeff#1#2{{\textstyle {\frac {#1}{#2}}}}

\def\half{\coeff 12}
\def\N{{\cal N}}
\usepackage{verbatim}   
\usepackage[all]{xy}
\usepackage{bm}  
\def\Dslash{{\rlap{\raise 1pt \hbox{$\>/$}}D}}
\def\Pslash{{\rlap{\raise  1pt \hbox{$\>/$}}\,\partial}}

\newcommand{\diff}{\mathrm{d}}

\newcommand{\Diff}{{\mathcal{D}}}

\newcommand{\be}{\begin{equation}}      
\newcommand{\ee}{\end{equation}}      
\newcommand{\bea}{\begin{eqnarray}}      
\newcommand{\eea}{\end{eqnarray}}

\newcommand{\im}{\mathrm{i}}

\newcommand{\rme}{\mathrm{e}}

\begin{document}


\title{TQFT at work for  IR-renormalons, resurgence and  Lefschetz decomposition}

\author{Mithat \"Unsal}
\email{unsal.mithat@gmail.com}
\affiliation{Department of Physics, North Carolina State University, Raleigh, NC 27695, USA}

\begin{abstract}
We investigate the implications of coupling a topological quantum field theory (TQFT) to  Yang-Mills theory with $SU(N)$ gauge group in the context of the IR-renormalon problem. 
Coupling a TQFT to QFT does not change the local dynamics and  perturbation theory, but it does change the bundle topology. 
 Crucially, the configurations with integer topological charge but  fractional action  contribute to the path integral of the original theory.  In the semi-classical regime,  these are critical points at infinity, called neutral bions,  and since ${\rm Arg}(g^2)=0$ is a Stokes line, their  Lefschetz thimbles are 
two-fold  ambiguous.  Therein, the ambiguity in the   gluon condensate  is sourced by the   neutral bions. 
 The Fourier decomposition of multi-branched observables at strong coupling is compatible  with the saddle decomposition at weak coupling.  TQFT coupling and non-renormalization of  $\theta$ angle impose constraints   and helps to identify  IR-renormalon. 
 In certain IR-CFTs, we prove irrelevance of bions, and absence of IR-renormalons.  
 
\end{abstract}


\maketitle



\noindent
{\bf Introduction:} In this work, we explore the connection between two important concepts 't Hooft introduced more than 40 years ago:
 the IR-renormalon problem  \cite{tHooft:1977xjm,   Beneke:1998ui,  David:1983gz, Shifman:2013uka, Dunne:2016nmc}  and  QFT in the  't Hooft flux background 
 \cite{tHooft:1979rtg,  
  tHooft:1981nnx}.  The latter is also called   coupling a TQFT to QFT 
\cite{Gukov:2013zka, Kapustin:2014gua, Gaiotto:2017yup}. 
   The IR-renormalon  problem is tied to the non-Borel summability of perturbation theory, and the question of a possible microscopic mechanism that 
 may fix the pathologies thereof.   
 The TQFT coupling changes only the sum over bundle topologies, while changing nothing locally in an arbitrarily large ball $B_d$ in a $d$-manifold $M_d$.  The two notions seem unrelated. 
    We claim that in a rather general class of theories, the TQFT coupling carries  crucial data to solve the IR-renormalon problem.  This may sound  
 strange  since  the TQFT coupling does  {\it nothing} locally, but it is exactly this  curious fact that  will be our guide.

  We consider  $SU(N)$ 
  pure   Yang-Mills theory  on  a 4-manifold $M_4=T^4,  \R^4   $ (and remark briefly   on the $\mathbb {CP}^{N-1}$ model  in 2d), and their supersymmetric completions, and an  IR-conformal theory. 
YM theory has a $\Z_N^{[1]}$ 1-form center-symmetry, for which we can turn on a background field or gauge.  The  gauged version is called the $PSU(N)_p $  theory, where $p=0, 1, \ldots, N-1$ is  a discrete  
label.  The $SU(N)$ and $PSU(N)_p$ theories are locally the same; 
 their   perturbation theory, non-Borel summability, and  IR-renormalon problems are  {\it exactly} the same.

On
the other hand, their sums over bundle topologies are different.  The topological sectors  in the $SU(N)$ theory  are classified according to   
instanton charge density (which integrates to an integer) \cite{Belavin:1975fg, Vainshtein:1981wh}, while the  
bundles  in the 
$PSU(N)_p$  theory are  much finer, being classified by two topological quantities: instanton charge density and 2nd Stiefel-Whitney class $ w_2 \in H^2(M_4, \Z_N)$.  The topological charge in the $PSU(N)_p$ theory  takes fractional values, $\frac{1}{N}\Z$, and in the BPS bound, the action of such configurations are also a fraction of an instanton action $S_I/N$. 
What should we make of this  different classical data? 

\vspace{0.3cm}
\noindent
{\bf Standard lore about perturbation theory on  $\R^4$:} 
Since perturbation theory is independent of the topological $\theta$ angle, so too is the ambiguity in the Borel resummation of the perturbation theory.  As a consequence, no topological configuration that carries $\theta$ angle dependence can be associated with a singularity in the Borel plane. 
The so called instanton singularity in the Borel plane  of $SU(N)$ theory  is  always  an instanton-antiinstanton  singularity \cite{Lipatov:1976ny,Bogomolny:1977ty}. 
The $[I \bar I]$ singularity in the Borel plane is located at:
\begin{align}
t_{ [I \bar I]}^{\R^4}  = 2S_I g^2 = 16 \pi^2
\end{align}
and the corresponding ambiguity is of order $ \pm i e^{-2S_I} $. 
This   is sourced  by  the factorial growth of the number of Feynman diagrams. 

However, there are other more important singularities in the Borel plane, much closer to the origin
\cite{tHooft:1977xjm, Beneke:1998ui}.
These arise  from the integration over low  and high momentum domains combined with the  singularity in the running of the coupling constant.    For example,  integration over low energy domain in the   Adler function  produces  a factorial growth  \cite{Beneke:1998ui}.
   \begin{align}
  D (Q^2)  \sim   \frac{\alpha_s}{2}\,\sum_{n=0}^\infty   
  \frac{n!}{  \left(  4S_I/ \beta_0 \right)^n   }
    \label{nineppp} 
  \end{align}
 Borel resummation of this series produces an ambiguity of order  $ \pm i e^{-4S_I/\beta_0} \sim \pm i \Lambda^4$, located at 
 \begin{align}
t_{ \rm ren.}^{\R^4} = \frac{4}{ \beta_0} S_I g^2  =  \frac{2}{ \beta_0} t_{ [I \bar I]}^{\R^4} = \frac{6}{11 N } t_{ [I \bar I]}, 
\label{borel1}
\end{align}
which is much closer to the  origin than $t_{ [I \bar I]}$ is.  
The renormalon on $\R^4$ is  believed to be fixed by an ambiguity in the gluon condensate,  ${\rm Im} \big \langle \frac{1}{N}  \tr F_{\mu \nu} F^{\mu \nu}  \big \rangle_{\pm}  \sim \pm \im   \Lambda^4  $, as argued in \cite{Parisi:1978az} based on OPE analysis of   \cite{Shifman:1978bx}.  There is also recent 
 lattice evidence  \cite{Bali:2014sja, Bauer:2011ws}.  However, a microscopic mechanism through which this take place in the strong coupling regime  is not  yet known.   For recent works on renormalons, see  \cite{Argyres:2012ka, Argyres:2012vv,  Dunne:2012ae,  Dunne:2012zk,  Shifman:2014fra,  Marino:2019eym, Marino:2019fvu, Marino:2021six,  Fujimori:2017oab, Fujimori:2018kqp,   Fujimori:2016ljw,Misumi:2016fno, Misumi:2014jua, Misumi:2019upg,   Pazarbasi:2019web}.

\vspace{0.3cm}
\noindent
{\bf Coupling YM to $\Z_N$ TQFT and  bundle topology:} 
We describe the path integrals of  $SU(N)$ and   $PSU(N)_p$ theories in regard to their bundle topologies, which are respectively
\begin{align}
\sum_{W \in \Z} \;  \int_{W} da  \qquad {\; \rm vs. \;}   \;\;  \sum_{w_2 \in H^2(M_4, \Z_N)}   \sum_{W \in \Z} \;  \int_{W, w_2} da
\end{align}
In the $SU(N)$ theory,  the integral  is over connections in a fixed bundle labelled by $W$ together with a sum  over  $W \in \Z$.   
In the  $PSU(N)_p$ theory,   the integral is  over connections in a fixed bundles labelled by   $W$ and  $w_2 $ together with a sum over $W\in \Z$ and $w_2 \in H^2(M_4, \Z_N)$. A fixed $w_2 \in H^2(M_4,\Z_N)$ evaluated on various 2-cycles gives the $\Z_N$ 't Hooft fluxes of the corresponding bundle.

Start with $SU(N)$ Yang-Mills theory  with action:
\be
S={1\over 2g^2} \int \tr[F\wedge \star F]+{\im\, \theta \over 8\pi^2}\int \tr[F\wedge F] 
\label{SU}
\ee
where the second term is the properly  quantized topological term, 
$\im \theta Q  \in  \im \theta   \Z$. 
This theory has a 1-form  $\Z_N^{[1]}$   symmetry. 

The   $\Z_N$ TQFT  can be defined as 
a path integral over 
a pair of properly quantized  fields $(B^{(2)}, B^{(1)})$  
with the insertion of the phase  $ {\rm exp}\left[  \im p {N\over 4\pi}\int B^{(2)}\wedge B^{(2)}   \right]   $. 
The TQFT  has 
 a 1-form gauge invariance. To couple  the $SU(N)$ YM theory to the background  $B^{(2)}$ field, we first define a $U(N)$   field  $\tilde a = a+{1\over N}B^{(1)}$. The  1-form gauge invariant combination  is  $\widetilde{F}-B^{(2)}$ where $\widetilde{F}$ is the  $U(N)$   field strength.   
The action of the $SU(N)$  theory in the TQFT  background is  \cite{Kapustin:2014gua}:
\begin{align}
{S}[ B^{(2)},  \widetilde a] = &{1\over 2g^2 }\int  \tr[  (\widetilde{F}-B^{(2)})\wedge \star  (\widetilde{F}-B^{(2)})]  & \cr
+ &  {\im\, \theta \over 8\pi^2}\int \tr[  (\widetilde{F}-B^{(2)}) \wedge  (\widetilde{F}-B^{(2)})]  &
\label{SUflux}
\end{align}
Summing over all  backgrounds gives the $PSU(N)_p$  theory: 
 \begin{align}
 Z_{PSU(N)_p} =  \int  \Diff   B^{(2)} \Diff B^{(1)}  \Diff  \widetilde a \;    \delta( N B^{(2)}-\diff B^{(1)})   &  \cr
 \;  \qquad  \rme^{\im p {N\over 4\pi}\int B^{(2)}\wedge B^{(2)}   } \;    \rme^{-S[ B^{(2)},  \widetilde a] }  \qquad  
 \label{top-5}
 \end{align} 
Using $B^{(2)} = \frac{2 \pi}{N} \ell_{\mu \nu}  \diff x_{\mu} \wedge   \diff x_\nu$,   $ \ell_{4i}= \ell_i, \ell_{ij}=\epsilon_{ijk} m_k $ to denote  
   't Hooft  fluxes,    the partition functions for  $SU(N)$, $SU(N)$ in $B^{(2)}$ background,  and $PSU(N)_p$  theories can be expressed as: 
\begin{align}
&Z_{SU(N)} =  \sum_{W \in \Z}        e^{ \im \theta W}  Z_{W}  &    \cr
&Z_{SU(N)} (\ell, m) =    \sum_{W \in \Z}       e^{ \im \theta \left( W+\frac{ (\ell \cdot m)}{N}  \right) } Z_{W}( \ell, m)  &    \cr
&Z_{PSU(N)_p} = 
  \sum_{ \substack{W \in \Z \\  \ell, m  \in (\Z_N)^3} }     e^{  \im  \frac { 2 \pi }{N} p  \; (\ell \cdot m)}      e^{ \im \theta \left( W+\frac{ (\ell \cdot m)}{N}  \right) } Z_{W}( \ell, m)      \qquad  \qquad &
\end{align}
The  gauging procedure admits the freedom to add a topological phase, a discrete topological theta angle, $\theta_p=  \frac{2 \pi  p }{N}$  to each network configuration of topological defects, hence the subscript in $PSU(N)_p$.    Here are three implications of this  discussion for the resurgence program \cite{Dunne:2016nmc}:

{\bf 1)} 
The self-dual saddles in  $PSU(N)_p$  theory are solutions of the self-duality equation:
\begin{align}
(\widetilde F-B^{(2)})=  \mp \star( \widetilde F-B^{(2)})
\label{Mod-inst}
\end{align}
The configurations that saturate the   BPS bound  have action:  \cite{tHooft:1979rtg, tHooft:1981nnx, vanBaal:1982ag, Gonzalez-Arroyo:2019wpu, GarciaPerez:1992fj,   GarciaPerez:1993jw, Unsal:2020yeh}
\be
S  =  \frac{8 \pi^2}{g^2}  \left| W+\frac{ (\ell \cdot m)}{N}   \right| \in  \frac{S_I}{N}  \Z^{\geq 0}
 \label{Mod-inst-ac}
\ee

 {\bf 2)} 
If we take a minimal BPS and anti-BPS   configuration in  $PSU(N)_p$,   we can construct  critical points at infinity   with $W=0$ and action $2S_I/N$
that can be lifted  to   $SU(N)$  bundle \cite{Unsal:2020yeh}. 
This is the first  point where  our construction is in sharp disagreement with the 
 widely accepted  perspective  \cite{ tHooft:1977xjm, Bogomolny:1977ty, Beneke:1998ui, Parisi:1978bj} 
which  asserts  that the leading non-BPS semi-classic  configuration in the $SU(N)$  theory must have an action  $2S_I$ 
and as such, semi-classical saddles 
must be irrelevant to  IR-renormalon problem.

{\it If} the theory is in a weakly coupled  domain \cite{Unsal:2012zj,   Argyres:2012ka},  
  $2S_I/N$ configurations  produce  a singularity in the Borel plane at  
\begin{align}
t^{*} = \frac{1}{ N} 2 S_I =  \frac{1}{ N } t_{ [I \bar I]}.
\label{bion}
\end{align}
One may be tempted to  think that despite  the proximity to the origin of the Borel plane by a factor of $N$ relative to $t_{ [I \bar I]}$, 
 the mismatch   between  \eqref{bion} and  \eqref{borel1}  
   is a clear failure, 
    and the singularity  \eqref{bion} is unrelated to the gluon condensate which is believed to be a macroscopic resolution of renormalon ambiguity in  YM theory on $\R^4$.  In fact,  there is some criticism of the semi-classical ideas in this regard, which we  also address in this paper  \cite {Anber:2014sda, Ishikawa:2019tnw, Ashie:2019cmy, Ishikawa:2019oga, Ishikawa:2020eht, Morikawa:2020agf, Ashie:2020bvw, Yamazaki:2019arj}

 {\bf 3)} An important step  toward  the resolution of this issue  is from somewhere unexpected:  the non-renormalization theorem for the  theta angle,  
which is  the  coefficient of the topological term. In $PSU(N)_p$ theory, $\theta \sim \theta+ 2 \pi N$ and in $SU(N)$, $\theta \sim  \theta+ 2 \pi$.   
However, in both cases, the  local observables are  $2 \pi$ periodic,  {\it multi-branched} functions, and each branch is $2 \pi N$ periodic 
\cite{Witten:1980sp, Witten:1998uka}, both in weak and strong coupling. The form of  theta angle dependence will provide  severe constraints.

 \vspace{0.3cm}
 \noindent
{\bf From QFT  to QM  on $\bm{ (N-1)}$-simplex with TQFT coupling:}
Consider  Yang-Mills theory in the following setting. First compactify the theory on $\R^4$ down to small $\R^3 \times S^1_L$ and insert a double-trace deformation on $S^1_L$ to stabilize the 0-form part of  center symmetry. The resulting theory satisfies adiabatic continuity: it  is continuously connected to pure YM  on $\R^4$ in the sense of all gauge invariant order parameters \cite{Unsal:2008ch, Shifman:2008ja}.  
This proposal is tested on the lattice, and it works \cite{ Bonati:2018rfg,  Bonati:2019kmf, Athenodorou:2020clr}. 
 Then,  further compactify the theory on  $ \R \times T^2 \times S^1_L$  where the $T^2$ size is of order $LN$ for reasons to be explained.      We map this system to  quantum mechanics within the Born-Oppenheimer (BO) approximation. Finally, when we study the partition function in QM, we compactify the Euclidean time direction, and study the theory on the 4-manifold 
\begin{align} 
 M_4 = S^1_\beta \times  T^2_{LN}  \times S^1_L, \;\;  L\Lambda \ll 1,  \; \beta \Lambda \gg 1 \;, 
 \label{4man}
 \end{align} 
   with all  TQFT backgrounds turned on.

Let us turn on a background  magnetic 't Hooft and GNO   flux \cite{tHooft:1979rtg, Goddard:1976qe} through the  the 12-plane: 
\begin{align}
  \frac{N}{2\pi}  \int_{\Sigma_{12}}  B^{(2)} = k \equiv \ell_{12}, \;\;\; \Phi= \int_{\Sigma_{12}}  \bm B = \frac{2 \pi} {g}  \bm \mu_k
\label{flux}
\end{align}
where  ${\bm \mu}_k \in \Gamma_{w}^{\vee}$ is an element of the co-weight lattice. 
  The energy of the  background flux configuration is \cite{Banks:2008tpa}:
  \begin{align}
E & =   \half  \int_{\Sigma_{12}}  {\bm B}^2=  \frac{\Phi^2}{2{\rm Area}(T^2)} 
\label{energy}
\end{align}
In both $SU(N)$  and $PSU(N)_p$ theory, the  dynamical monopoles  are  elements  of the co-root lattice,  ${\bm \alpha} \in \Gamma_{r}^{\vee}$.  

\begin{figure}[t]
\begin{center}
\includegraphics[width = 0.50 \textwidth]{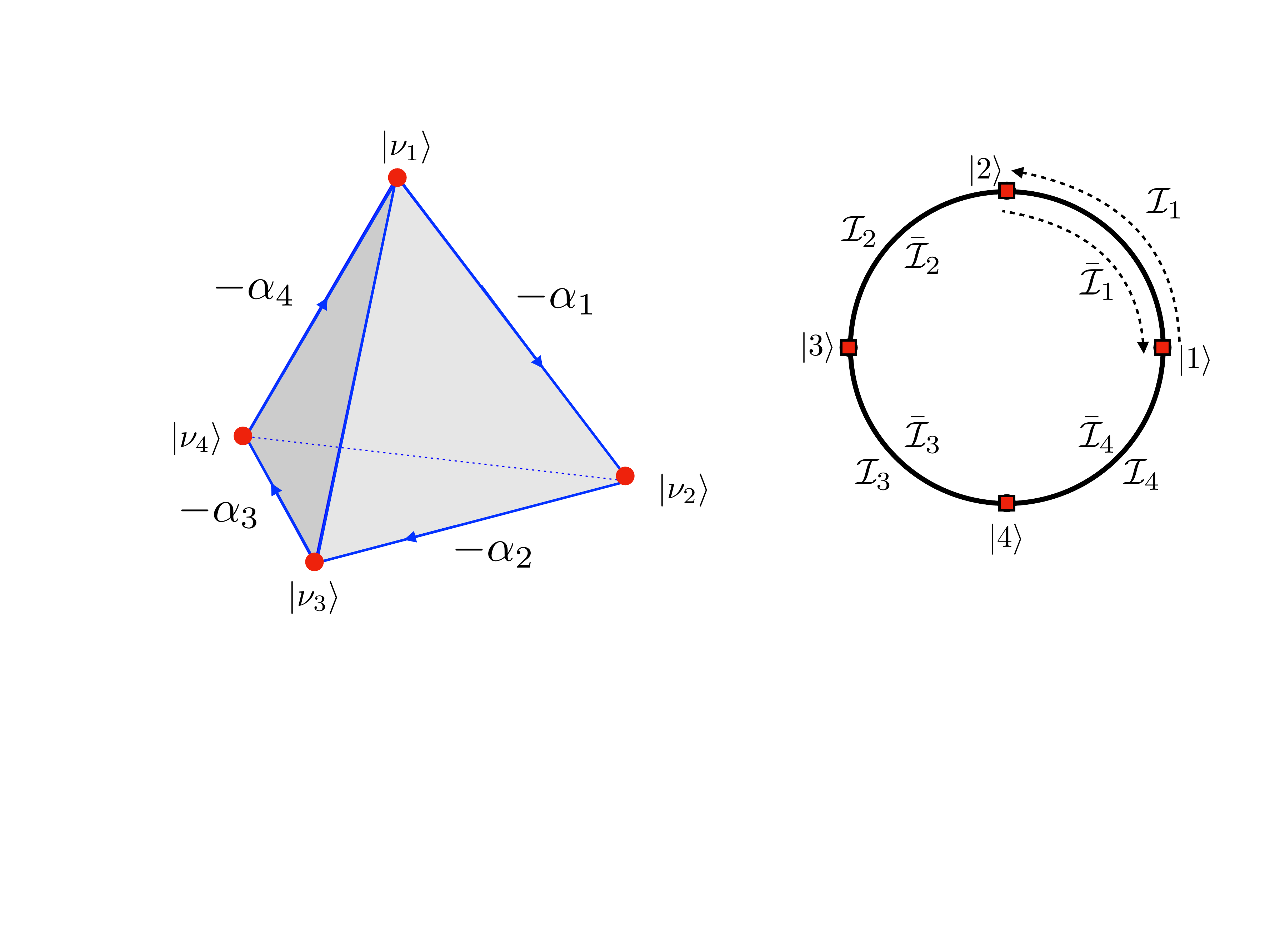}
\vspace{-2.5cm}
\caption{In the TQFT background,  and within Born-Oppenheimer approximation, the $SU(N)$ YM theory reduce to a quantum mechanical system of a particle on a $(N-1)$-simplex. $S_N$ gauge structure is Higgsed, but 0-form part of  $\Z_N$  center acts as a cyclic permutation. Picture is for $N=4$.  
 }
 \vspace{-0.7cm}
\label{fig:torusqm}
\end{center}
\end{figure}

Take $k=1$.   First, we note that there are two  classes  of tunneling events \cite{Unsal:2020yeh}. Since the energy depends  on the area of $T^2$  inversely as in 
\eqref{energy}, one type of tunneling is among the states   which become degenerate only in ${\rm Area}(T^2) \rightarrow \infty  $ limit (see  \cite{Banks:2008tpa}, page 226). 
However,  in  the  flux background  \eqref{flux}, there exist  $N$ {\it exactly degenerate} minima     even at {\it finite}  ${\rm Area}(T^2)$, given by:
 \begin{align}
| {\bm \nu}_a \rangle  \equiv  | {\bm  \mu}_1 -    {\bm  \alpha}_1 \ldots -   {\bm  \alpha}_{a-1} \rangle   \qquad a=1, \ldots,  N
     \label{degenerate-states}
\end{align} 
These states can be visualized  as  the $N$  vertices of an $ (N-1)$-simplex.  Despite the fact that these points form an $(N-1)$-simplex, the system does not have the $S_N$  permutation symmetry. Only the tunneling between  adjacent pairs  
are minimal   and all others 
 are non-minimal action. 
  The $\Z_N^{[0]}$ center-symmetry acting on the Polyakov loop along $S^1_L$ cyclicly permutes these  tunneling events.   The minimal tunneling events form  a  {\it closed loop}    in field space:
 \begin{align}
 | {\bm \nu}_1 \rangle \underbrace {\longrightarrow}_{    -{\bm  \alpha}_1}   | {\bm \nu}_2 \rangle   \underbrace {\longrightarrow}_{    -{\bm  \alpha}_2}  
 \cdots 
  \cdots   
    \underbrace {\longrightarrow}_{    -{\bm  \alpha}_{N-1}}    | {\bm \nu}_N \rangle \underbrace {\longrightarrow}_{    -{\bm  \alpha}_N} | {\bm \nu}_1 \rangle
       \label{tunnelilngsQM}
\end{align} 
Note that  the vacuum of YM theory on $M_4$ with  flux \eqref{flux}  is identical to the ${\mathbb {CP}}^{N-1}$ model on $\R \times S^1_L$ with twisted boundary condition \cite{Dunne:2012ae,  Dunne:2012zk,  Yamazaki:2017ulc, Yamazaki:2017ulc, Yamazaki:2017dra}.  
 If we apply the BO approximation a second time, see Fig.~\ref{fig:torusqm},   
 Yang-Mills theory  maps to a particle on a circle with $N$ mimima. 
In the 
 latter, Ref.~\cite{Sueishi:2021xti}, generalizing  \cite{pham},   proved all-orders   resurgent cancellations 
   by using exact WKB.  
    What is the implication of this  all-orders resurgent cancellation    for IR-renormalon problem?
   
 \vspace{0.3cm}
\noindent
{\bf Gluon condensate from small to large-$L$:}
We can first calculate the gluon condensate on  $T^3_{\rm large} \times S^1_L$ at small-$L$ by weak coupling methods. Then, by using dimensional analysis and multi-branch structure of theta angle dependence,  we show compatibility with   large   $T^4_{\rm large}$ limit.  The  hero of 
 this story is non-renormalization of theta, and multi-branched structure of the condensate. 

In  the first two orders in semi-classics,  we can consider the Euclidean description  of the  vacuum  of the quantum mechanical system as a dilute gas of  fractional instantons and bions.  There are    $N$ types of fractional-instantons  ${\cal M}_i$ with topological charge $Q=1/N$, and  complex fugacity     $e^{-S_I/N  + i \theta/N}$. 
  The  fractional topological charge follows from    ${N\over 8\pi^2}\int B^{(2)}\wedge B^{(2)}  \in \frac{1}{N} \Z$
which is valid  {\it both} at  small,   and    large $M_4$ where semi-classics is no longer valid  \cite{Unsal:2020yeh}.
There are also  $N$ types of neutral bions  ${\cal B}_{i, \pm}= [{\cal M}_i \overline{\cal M}_i]_{\pm}$ with zero topological charge, but action $2S_I/N$.    These configurations in quantum mechanics are  attached to critical points at infinity \cite{Behtash:2018voa, Nekrasov:2018pqq}. Integration over the  quasi-zero mode Lefschetz thimble produces a two-fold ambiguous result, whose imaginary part is   $\sim \pm i e^{-2 S_I/N } $.  
 What do we make of this  at strong coupling?

Due to the trace anomaly, the gluon condensate is proportional to minus vacuum energy density.  Vacuum energy  for QM  shown in Fig.~\ref{fig:torusqm}  is easiest  to diagonalize using the  tight binding Hamiltonian.  Incorporating first and second order effects in semi-classics, we find:
\begin{align}
\Big \langle \textstyle \frac{1}{N}\tr F^2 \Big \rangle_{\pm}  (\theta)
 &=   \max\limits_{ q}   \left[   c_1 \Lambda^{\frac{11}{3}}   {\cal R}^{- \frac{1}{3}}  \cos\Big(\frac{2\pi q+\theta}{N}\Big)  \right.  \cr 
& \left. +  c_2 \Lambda^{\frac{22}{3}}   {\cal R}^{\frac{10}{3}}  \cos2 \Big(\frac{2\pi q+\theta}{N}\Big)   + \ldots  \right]
\cr   & + c_3  \Lambda^{\frac{22}{3}}  
{\cal R}^{\frac{10}{3}}  \pm i \pi   c_4 \Lambda^{\frac{22}{3}}  {\cal R}^{ \frac{10}{3}} + \ldots  
\label{condensate}
\end{align} 
where ${\cal R} \equiv {\cal R} (LN)$ is the the fractional instanton size function, proportional to   $LN \sim m_W^{-1}  $ in the semi-classical regime.  
On  large  $T_4$, 
 the condensate  is   expected to be:
\begin{align} 
 \Big \langle { \textstyle \frac{1}{N}} \tr F^2  \Big   \rangle_{ \pm }  (\theta)  =  \max\limits_{ q}    \Lambda^4  h\Big( \frac{\theta + 2 \pi q}{N} \Big)     \pm i a \Lambda^4\; , 
 \label{largeM4}
\end{align} 
where $h (\theta/N)$ is  
$2 \pi N$ periodic function \cite{Witten:1980sp, Witten:1998uka}. 
Here comes  the   important part of  the story.

\noindent
{\bf  Fourier  vs.  Lefschetz decomposition:} The appearance of   $\rme^{ \im \frac{\theta}{N}  k}, k \in \Z $   in \eqref{condensate} arises from calculable semi-classics, both on $M_4$ \eqref{4man} and $\R^3 \times S^1$. This is the weak coupling regime. The function that appears in \eqref{largeM4}
 $h (\theta/N)$ can be expressed in terms of a complete  Fourier basis: 
 \begin{align}
  \{ {\rm Span}(\rme^{ \im \frac{\theta}{N}  k}),  \;  k \in \Z \}\;.  
  \label{basis}
  \end{align}
 The decomposition into  $\rme^{ \im \frac{\theta}{N}  k}, k \in \Z $  is  inevitable, either on {\it small or large}  $M_4$  as a consequence of  
  thinking $SU(N)$ theory in TQFT backgrounds.  
  All  local non-perturbative  observables  must  be multi-branched   where each branch can be decomposed  in the basis \eqref{basis}.

   $\rme^{  \pm \im \frac{\theta}{N}  } $  is minimally  accompanied with  
$ \rme^{-S_I/N}$ in weak coupling, and 
 the expansion organizes itself in positive integer  powers of   $\Lambda^{11/3}$. 
At second order in semi-classics,   an  important effect is due to   neutral bions. 
 Since ${\rm Arg}({g^2})=0$ is a Stokes line,  the neutral bion amplitude is two-fold ambiguous    \cite{Behtash:2018voa}.  This is of order 
  $ \pm \im e^{-2S_I/N}  \sim  \pm \im \Lambda^{22/3}$ and is the configuration that enters to all orders  resurgent cancellations \cite{Sueishi:2021xti}    in  QM limit.  

 However,  the expected ambiguity in the condensate  on  $\R^4$ is  of the form $\pm \im e^{-4S_I/ \beta_0}$, corresponding to $\pm  \im \Lambda^4$. Wouldn't it be nicer if semi-classics produced   $\Lambda^4$ factors as in \eqref{largeM4}? The answer is,  no.  The only terms allowed in the semi-classical expansion compatible with $PSU(N)_p$ bundle are of the form (set $p=0$ for convenience):
 \begin{align} 
 \left\{ \Lambda^{\frac{11}{3} (n+ \bar n)}      {\cal R}^{\frac{11}{3} (n+ \bar n) -4}     \;   \rme^{ \im \frac{\theta}{N} (n- \bar n)} \right \}
   \label{series}
 \end{align} 
 where $n \geq 0, \bar n \geq 0, 
 (n, \bar n) \neq (0, 0) $. 
 Based on this, it is clear that   if we were to obtain $\Lambda^4$ 
  in semiclassical domain of $PSU(N)_0$ theory, 
  it  would accompany  $e^{i \frac{12}{11} \frac{\theta}{N}}$ and   be a  \emph {contradiction} with  $2 \pi N$ periodicity.
   Therefore, it is  not only futile,   it is  in fact wrong  to look for a saddle which would produce $\pm \im e^{-4S_I/ \beta_0}$, answering a concern in 
   \cite{Yamazaki:2019arj}. 
 The  $\theta$-dependence of the semi-classical expansion \eqref{series}  
  can also be viewed as a consequence of bundle topology in $PSU(N)_p$ theory. 
Dimensional analysis tells us that the monomials  \eqref{series}  must be accompanied with  powers  ${\cal R}\sim LN$ to match the scaling dimension of the condensate.   However, the general form of  ${\cal R}(LN)$-function  is unknown.

  The completeness of the Fourier  basis   and its matching with the Lefschetz thimble   decomposition of the partition function in terms of  
 critical points (including the ones  at infinity)    gives us a realistic hope that the Lefschetz  decomposition of path integral  may well  be complete!    (See also \cite{Dunne:2012ae, Krichever:2020tgp, Gukov:2016njj,   Witten:2010cx, Kontsevich} for similar standpoint in different theories.)
The Lefschetz decomposition  is  exact in QM (which is a limit of YM on  $ T^3 \times \R$)  as per translation of exact WKB result to path integral \cite{Sueishi:2021xti}, and also 
  produce the correct non-perturbative dynamics at small $S^1_L \times \R^3$  as reviewed in 
  \cite{Dunne:2016nmc}.   The real difficulty is to understand the implications of this decomposition at  strong coupling.

What happens as  $\Lambda L N $ gets larger? 
In the $\Lambda L N \gg  1$ regime,  the theory  must gradually become a  one-scale problem, dependent only on   $\Lambda$, and the 
 condensate must saturate to its value on  $\R^4$.  In the large-$N$ limit, this fact is guaranteed by large-$N$ volume independence, which is proven by using  lattice field theory \cite{Eguchi:1982nm, GonzalezArroyo:1982hz, Unsal:2008ch}. 
 The semiclassical BO-approximation  used to derive  \eqref{series} is valid provided $\Lambda L N \lesssim 1$.

Although semi-classical effective field theory is  not sufficient to address  what happens as the theory moves from the semi-classical regime \eqref{condensate}  to   the strong coupling regime \eqref{largeM4},   there is a sense in which it can be useful.  
On $T^3_{\rm large} \times S^1_L$ emulating $\R^3 \times S^1_L$, 
  the Debye length   is $\xi \equiv m_g^{-1}  \sim  \Lambda^{-1} (\Lambda  {\cal R})^{-5/6}$  due to the proliferation of the  fractional instantons  \cite{Unsal:2008ch}. The size of the fractional instantons in weak coupling  is  $  {\cal R}   \sim  L N  $.  Polyakov argues that (see \cite{Polyakov:1987ez}  page 91),  if the finite correlation length cuts off the growth of the fractional instanton size, the EFT pushed to the boundary of its region of validity   can provide a {\it qualitatively} accurate description.  We  are in search of  a qualitative description after all, and  take this  as our   assumption. Given the assumption, 
   the condensate  saturates  to expected behavior on $\R^4$  
  \begin{align} 
 \Big \langle { \textstyle \frac{1}{N}} \tr F^2  \Big   \rangle_{ \pm }  (\theta)  =  \max\limits_{ q}     \Lambda^4  \Big( \sum_{k \in \Z} c_k  e^{ i \frac{\theta + 2 \pi q}{N}  k } \Big)     \pm i a  \Lambda^4 
 \label{largeM4-2}
\end{align} 
At this stage, the ambiguities generated by neutral bions 
 \eqref{bion}
source and transmute to    the ambiguities that are associated 
with condensates \eqref{borel1}, and location of leading singularity in Borel plane flows as:
\begin{align}
 t_{\rm bion }^{\R^3 \times S^1}=  \frac{2S_I}{ N }  \;  
 \Longrightarrow \;  t_{ \rm ren.}^{\R^4} =  \frac{4 S_I}{ \beta_0} 
\end{align}
 In fact,  neutral bions  provide a match to  all IR-renormalon singularities located at   $t_{ \rm ren.}^{\R^4} =  \frac{S_I}{ \beta_0} (4+2a), a=0,1,2, \ldots$.   The non-trivial aspect of  \eqref{largeM4-2} is that Lefschetz decomposition of saddles 
 extrapolated at the boundary of EFT  matches  Fourier decomposition of non-trivial multi-branched functions at strong coupling!
 This perspective provides a  microscopic  definition of IR-renormalon.

    \noindent
{\it  Definition:} 
The IR-renormalons in 4d gauge theories  are composites of  (semi-classical)  fractional  instantons configuration in the $PSU(N)_p$  bundle that lifts to $SU(N)$ bundle,  and that are   dressed by strong dynamics.  The effect of dressing is to  saturate the  $ {\cal R}$-function around the scale $\Lambda^{-1}$,   if the running coupling   develops a singularity (pole, branch point)  thereof. 

 As far as we can see, the saturation of   $ {\cal R}$-function \cite{Polyakov:1987ez}  is a   unique way to make {\it 1)} theta angle periodicities  {\it 2)} multi-branched structure  \cite{Witten:1980sp, Witten:1998uka},  {\it 3)} weak coupling analysis on $\R^3 \times S^1$   \cite{Argyres:2012vv,  Dunne:2012ae}  and  {\it 4)} strong coupling expectation  on $\R^4$ \cite{tHooft:1977xjm,   Beneke:1998ui}  simultaneously consistent.

 \vspace{0.3cm}
\noindent
{\bf Examples:} We can present three  classes of  examples  that shows that  this construction  is correct more broadly.   
 
Coupling the 2d $\mathbb{CP}^{N-1}$  model 
 to a $\Z_N$ TQFT, the gauge field is classified   according to  both instanton density  and  $ w_2 \in   H^2(M_2, \Z_N)$; hence, topological charge  is quantized in units of $\frac{1}{N}\Z$. 
 On $\R^2$,  the  ambiguity in the spin-wave condensate is given by  ${\rm Im}  \langle \diff \bar z  \diff  z  \rangle_{\pm} \sim  \pm \im \mu^2 e^{-2S_I/\beta_0}=\pm \im \Lambda^2$  \cite{David:1983gz}. A pair of minimal BPS and anti-BPS  $U(1)/\Z_N$  configurations that can be lifted  to 
  $U(1)$ induces an   ambiguity 
  $ \pm \im \mu^2 e^{-2S_I/N}=\pm \im \Lambda^2 $  due to the fact that  $\beta_0= N$. This is  the  IR-renormalon.

  If we reduce 4d  $\N=1$ $SU(N)$  SYM theory and 2d   $\N=(2,2)$   $\mathbb{CP}^{N-1}$  models down   to QM with the   $B^{(2)}$ background,  both  reduce to  extended supersymmetric ${\cal N}=2$ QM.
 In ${\cal N}=2$ QM,  there are two types of   neutral bions,  both of which are   ambiguity free,  and the vacuum energy vanishes due to a relative hidden topological angle between  two types of  saddles \cite{Behtash:2015kva}, consistent with    the  vanishing of  condensates   (and their ambiguites) 
  \cite{Dunne:2015eoa}. 
  
In QCD with $n_{\rm adj}^W= 5$ adjoint Weyl  and  $N_{\rm f}= (\half  -\epsilon)N$  fundamental Dirac fermions,  which can also be coupled to a 
$\Z_N$ TQFT  \cite{Anber:2019nze}, 
 the ratio $ \xi/m_g \sim e^{8\pi^2/g_*^2N} \gg 1 $ as $LN \rightarrow \infty$, where $g_*^2N \sim \epsilon \rightarrow 0$  is the value of the coupling in a  Banks-Zaks  type   IR-fixed point \cite{Banks:1981nn}. Therefore, semi-classics is valid at {\it any} size $S^1_L \times \R^3$. 
The fractional instantons and bions contribution becomes irrelevant as $LN \rightarrow \infty$ as $  (LN)^{-4} e^{-8\pi^2/ \epsilon} \rightarrow 0 $.
The condensate  vanishes in  $\R^4$ limit  \cite{Dunne:2015eoa}, 
providing a semi-classical proof for the   absence of IR-renormalon  in IR-CFTs.

 \vspace{0.3cm}
 \noindent
{\bf Summary:} The fixing of the  Borel resummation ambiguities in $\R^d$ via the condensates and the fixing 
of it in small $M_d$ or $\R^{d-1} \times S^1_L$   via neutral bions are {\it not} different mechanisms. 
Since the gluon and spin-wave condensate are not  protected quantities,  
 they  have   $L$ dependence in the weak coupling domain. But the crucial point is that the non-trivial  theta angle dependence  of basis functions for Lefschetz decomposition of semi-classics and Fourier decomposition of the expected multi-branched strong coupling result match each other, and does not change with $L$.  
	Our construction shows  both the presence of   IR-renormalons, along its flow from semi-classical result   $ \frac{1}{N} t_{ [I \bar I]}$ to  strong coupling  result  	 $ \frac{6}{11N}  t_{ [I \bar I]} $ in Yang-Mills theory.  It also  proves the absence of IR-renormalon    in an  IR-CFT in decompactification limit.  
	We believe this reconciles the traditional perspective   \cite{tHooft:1977xjm,   Beneke:1998ui} with the modern semi-classical/TQFT perspective    based on neutral bions \cite{Argyres:2012ka,  Dunne:2012ae}.

\vspace{0.3cm}
\noindent
{\bf Acknowledgments.}  We are grateful to 
Misha Shifman,    Yuya Tanizaki, Aleksey Cherman, Sergei Gukov and Mendel Nguyen    for discussions. 
M.\"U. acknowledges support from U.S. Department of Energy, Office of Science, Office of Nuclear Physics under Award Number DE-FG02-03ER41260.


\bibliography{QFT-Mithat}

\bibliographystyle{JHEP}

%

\end{document}